\documentclass{article}

\usepackage{PRIMEarxiv}
\usepackage[utf8]{inputenc} %
\usepackage[T1]{fontenc}    %
\usepackage{hyperref}       %
\usepackage{url}            %
\usepackage{booktabs}       %
\usepackage{amsfonts}       %
\usepackage{nicefrac}       %
\usepackage{microtype}      %
\usepackage{lipsum}
\usepackage{fancyhdr}       %
\usepackage{graphicx}       %
\usepackage{subcaption}
\usepackage{caption}       %
\graphicspath{{figures/}}     %
\usepackage[table,xcdraw]{xcolor}
\usepackage{array}
\usepackage{colortbl}
\usepackage{placeins}
\usepackage{comment}
\usepackage{soul,color}
\usepackage{threeparttable} %
\usepackage{listofitems} %
\title{
Out-of-distribution materials property prediction using adversarial learning based fine-tuning
\thanks{\textit{\underline{Citation1}}: 
\textbf{Li et al.Improving OOD materials property prediction. .... DOI:000000/11111.}} 
}
\author{
  Qinyang Li \\
 Department of Computer Science and Engineering\\
  University of South Carolina\\
  Columbia, SC 29201 \\
   \And
   Nicholas Miklaucic \\
 Department of Computer Science and Engineering\\
  University of South Carolina\\
  Columbia, SC 29201 \\
  \And
 Jianjun Hu *\\
 Department of Computer Science and Engineering\\
  University of South Carolina\\
  Columbia, SC 29201 \\
  \texttt{jianjunh@cse.sc.edu} \\
}

\begin{document}
\maketitle
\begin{abstract}
The accurate prediction of material properties is crucial in a wide range of scientific and engineering disciplines. Machine learning (ML) has advanced the state of the art in this field, enabling scientists to discover novel materials and design materials with specific desired properties. However, one major challenge that persists in material property prediction is the generalization of models to out-of-distribution (OOD) samples, \textit{i.e.}, samples that differ significantly from those encountered during training. In this paper, we explore the application of advancements in OOD learning approaches to enhance the robustness and reliability of material property prediction models. We propose and apply the Crystal Adversarial Learning (CAL) algorithm for OOD materials property prediction, which generates synthetic data during training to bias the training towards those samples with high prediction uncertainty. We further propose an adversarial learning-based targeted fine-tuning approach to make the model adapt to a particular OOD dataset, as an alternative to traditional fine-tuning. Our experiments demonstrate the success of our CAL algorithm with its high effectiveness in ML with limited samples, which commonly occurs in materials science. Our work represents a promising direction toward better OOD learning and materials property prediction.
\end{abstract}
\section{Introduction}
The accurate characterization and prediction of material properties are pivotal for various applications, ranging from renewable energy systems and drug design to electronic devices and aerospace engineering. Traditionally, material property determination has relied on costly, time-consuming experimental methods. However, with the advent of machine learning, particularly deep learning, it has become feasible to develop predictive models that can efficiently analyze vast amounts of material data and provide accurate property predictions. For over a century, materials scientists have accumulated a large number of crystal structures through laborious experimentation, leading to modern crystal structure databases such as Inorganic Crystal Structure Database (ICSD) \cite{zagorac2019recent,allmann2007introduction,belsky2002new}, containing over 200,000 structures identified since 1913 with about 7,000 structures added each year, and more open databases such as the Materials Project \cite{jain2013commentary}, containing over 120,000 inorganic materials along with their properties. Despite the successes achieved by machine learning models for material property prediction \cite{wang2021compositionally,de2021materials,chen2019graph,xie2018crystal,jha2019irnet}, these models often struggle to generalize from their training data to test data that may be \textit{out of distribution (OOD)}—drawn from areas of the feature space dissimilar to the training data. These OOD predictions can be seriously inaccurate, leading to unreliable results and potentially hindering the development of novel materials \cite{hu2024realistic}.

Current state-of-the-art machine learning models in materials science operate under the classical paradigm, in which data is assumed to be independently, identically distributed (i.i.d) under a single data distribution. However, materials design problems often violate this core assumption. For example, material property optimization—finding materials that maximize or minimize a property—almost by definition involves predictions on new inputs that are not typical of the training dataset. Although there is some research conducted on molecular out-of-distribution (MOOD) problems \cite{RealWorldMolecularOOD,lee2023MOOD,yang2022learning}, there are only a few studies focusing on the crystal material OOD property prediction problem  \cite{hu2024realistic,omee2024structure}, in which the same challenge arises: training and testing data differ greatly in real-world deployment\cite{RealWorldMolecularOOD,yang2022learning}. Existing ML models usually have poor performance for such OOD inputs \cite{xiong2020evaluating,loftis2020lattice,back2024accelerated}. Direct optimization using a trained ML surrogate, even for standard properties like formation energy, often leads to exploiting inaccuracies in the model instead of correctly discovering new stable materials\cite{qi2023latent}. Even with techniques such as fine-tuning, traditional ML models will often either perform poorly or over-fit toward the target dataset. Meanwhile, the quality of the OOD dataset also plays a vital role in enhancing the performance of the model, as Zou \textit{et al.} discovered in their OOD research \cite{Zou_2023_ICCV}: an OOD test set with an inappropriate difficulty level causes poor performance. To address this issue, they used the adaptive calibrator ensemble (ACE) algorithm which combines the outputs of two calibrators trained on datasets with extreme difficulties. In our work we adopt the partial sampling algorithm\cite{magar2023learning} to actively adjust the difficulties of a portion of our training set. Another related work is Qi \textit{et al.} \cite{qi2023latent} which focused on training conservative objective energy models for gradient-based optimization for material structure prediction. Although we share the similar idea of utilizing gradients, their work is more towards exploring the search space with a latent space conservative model aimed to obtain the optimized crystal structure. Other related works in the MOOD field including Lee \textit{et al.} \cite{lee2023MOOD} are generative models that seek new materials in a targeted OOD region. 

The OOD material property prediction problem needs specifically designed OOD datasets for evaluation. Here we introduce such OOD datasets for our experiments, curated from a larger training set. We define the 145 composition-based physical attributes as the input $X$ and the target property formation energy as $y$ for all samples extracted from the Materials Project dataset\cite{jain2013commentary}. Our crafted OOD datasets can be categorized into three different classes: \emph{Covariate Shift} ($X$ shift), \emph{Prior Shift} ($Y$ shift), and \emph{Relation Shift} ($P(Y|X)$ shift)\cite{moreno2012unifying}. OOD samples pose great challenges for current i.i.d. based machine learning models. Table \ref{tab:SoArt_OOD} shows the high variation in the performance of several state-of-the-art composition-based neural network models for material property prediction, emphasizing the pressing need to address OOD prediction challenges. In this work, we propose the crystal adversarial learning (CAL) pipeline to address the OOD material property prediction problem based on the stable adversarial learning algorithm (SAL) as proposed in \cite{liu2021stable}.The main idea involves constructing an uncertainty set by differentiating features based on the stability of their correlations with the target. It generates adversarial samples by perturbing the unstable features using network gradients and optimizes seed sample weights through iterative gradient descent. Consequently, our CAL algorithm effectively identifies high correlation features that can be used to screen challenging OOD samples from the seed training samples for adversarial learning, expecting the model to achieve better performance for future real world OOD samples. At the same time, we utilize partial sampling techniques that iteratively select OOD samples from the training set that have high training losses. The combination of the two methods are shown to be able to improve our CAL model's OOD performance. CAL is also able to target specific OOD datasets and generate adversarial samples based on them. This is similar to the fine tuning technique used in conventional machine learning (ML). This targeted approach allows us to investigate how all three types of distribution shifts affect the model performance. By understanding these relationships, we aim to develop more robust and generalizable predictive models for material discovery, especially for special materials such as piezoelectric crystals and special properties such as phonon peaks.

Our work makes three key contributions:
\begin{itemize}
    \item We identify the critical need to address the limitations of existing machine learning models for OOD material property prediction.
    \item We propose CAL, a novel approach for OOD material property prediction that integrates adversarial learning with training set ranking. This approach enhances model robustness under diverse test scenarios marked by different input and target distributions.
    \item We demonstrate that our model can achieve better performance for low-data problems with distribution shift compared to competitive baselines.
\end{itemize}

\section{Method}\label{sec:headings}
\subsection{OOD Datasets}
We focus on the OOD material property prediction problem and use the formation energy and band gap as the examples for our main experiments. Using Matminer \cite{ward2018matminer}, we selected 84,190 materials from the Materials Project database\cite{jain2013commentary}. For each sample, we generated 145 composition-based physical attributes \cite{ward2016general}. We call the input information $X$ and the regression target $y$. $X$ is a set of statistics describing a material's composition, and $y$ is a property of that material—formation energy or band gap. There are three kinds of potential distribution shifts: \emph{Covariate shift}, \emph{Prior shift}, and \emph{Relation shift} \cite{moreno2012unifying}. Covariate shift is when the distribution of input data $X$ shifts between training and inference. Prior shift, also known as \textit{label shift}, occurs when the distribution of $y$ shifts from training to inference without a corresponding change in $X$. Relation shift, known by many other names including \textit{concept shift}, indicates a change in the relationship between the joint distribution of $X$ and $y$ without a shift in the individual marginal distributions of $X$ or $y$. In practice, real-world problems exhibit all three shifts to some degree, and these categories are not absolute. In our previous work\cite{li2023global}, we used the global mapping to show the relationship between $X$ and $y$. We consider the distribution of outputs $y$ as conditional on the inputs $X$. In this notation, covariate shift can be thought of as shift in $P(X)$, prior shift as shift in $P(y)$, and relation shift as shift in $P(y|X)$.

\begin{figure}[h!]
    \centering
    \begin{subfigure}[b]{0.49\textwidth}
        \centering
        \includegraphics[width=1.1\linewidth]{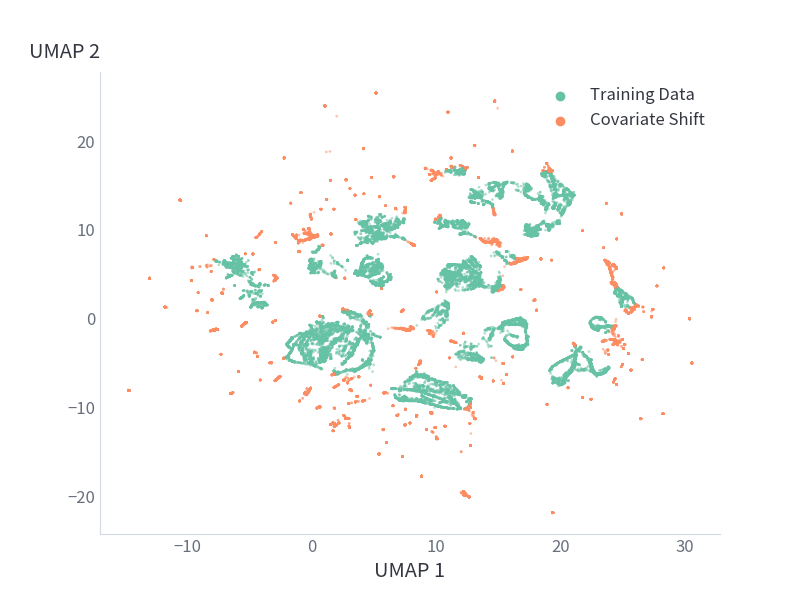}
        \caption{Covariate shift samples}
    \end{subfigure}%
    \begin{subfigure}[b]{0.49\textwidth}
        \centering
        \includegraphics[width=1.1\linewidth]{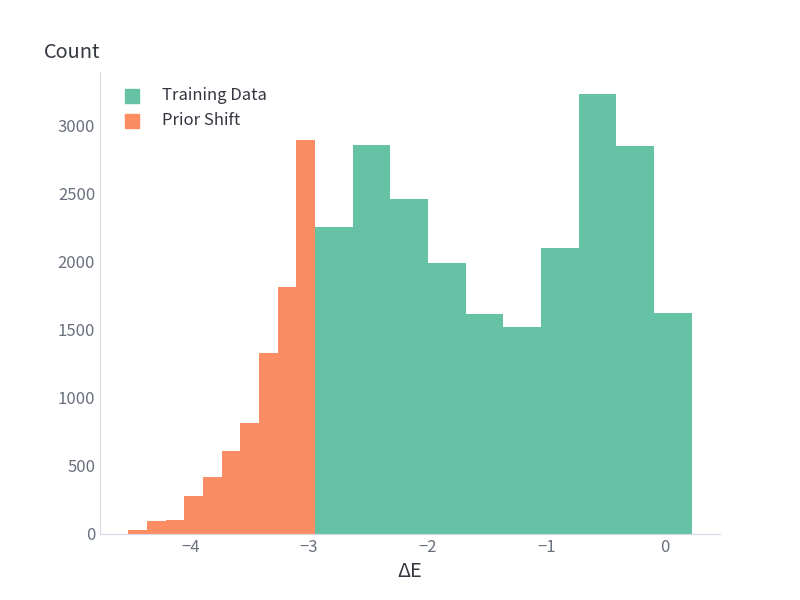}
        \caption{Prior shift dataset (for formation energy)}
    \end{subfigure}%
    \caption{Dataset selections for covariate and prior shifts. The covariate shift samples are selected from the periphery of the low-dimensional surface formed by the data. The prior shift dataset consists of outliers in formation energy.}
    \label{fig:data_shift}
\end{figure}

 We constructed several datasets to exhibit these different OOD challenges. For covariate shift, we use a low-dimensional representation of the data with UMAP \cite{umap}, a dimension reduction technique and select data points on the outside of the low-dimensional manifold (Figure \ref{fig:data_shift} a). For prior shift, we select the extreme values of $y$ (Figure \ref{fig:data_shift} b). For the relation shift set, we introduce a dataset for piezoelectric materials. Piezoelectric materials serve as a compelling case study because of their rarity and exceptional chemistry. There are currently no established quantitative models for piezoelectric modulus. The currently available data, which solely originates from experiments, is not readily amenable to traditional machine learning techniques due to its inherent variability and potential noise. That same uncertainty, however, makes machine learning models even more important for research and further exploration.

 To account for variations in composition features, we initially normalize our data. Then, as mentioned above, the OOD datasets can be categorized into three different classes: Covariate Shift ($X$ shift), Prior Probability Shift ($Y$ shift), and Relation shift ($P(Y|X)$ shift).

For the prior shift dataset, the existing dataset is first sorted by the property values $y$. Rather than employing a random split as is usually done in conventional ML studies, we partition the dataset based on a defined percentage, such as the top 20\% of $y$ values. This procedure results in a prior shift dataset where the target ($y$) of the test set remains unseen by the network. Figure \ref{fig:data_shift}(b) illustrates the generation of this supervised prior shift set for the $y$ values of the formation energy.

Constructing our covariate shift dataset is more involved because our input is 145-dimensional and cannot be sorted. Instead, we use UMAP \cite{umap} to reduce our data to 2 dimensions and then apply a kernel density estimate to select the points farthest from the main data distribution based on a defined percentage. The results and the embedding are shown in Figure \ref{fig:data_shift}(a), where the orange data points represent the selected out-of-distribution (OOD) samples.

To investigate relation shift ($P(y|X)$ shift), we opted for piezoelectric materials. This choice is motivated by two key factors. First, piezoelectric materials exhibit a unique relationship between their features ($X$) and target properties ($y$) compared to the majority of materials in the ICSD dataset. This distinct characteristic allows us to effectively explore how changes in this relationship impact model performance. Second, the availability of piezoelectric material samples is limited, reflecting a common bottleneck encountered in real-world materials science research. No quantitative model is currently available for piezoelectric modulus, indicating a need for further research. This limited data scenario provides a valuable opportunity to test our model's capability in handling situations with scarce samples.

\begin{figure}[ht]
    \centering
    \includegraphics[width=0.9\linewidth]{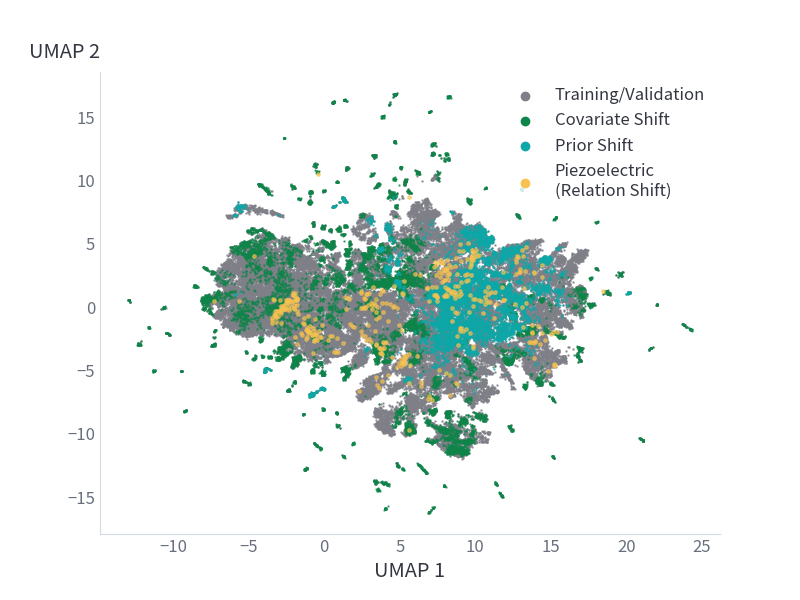}
    \caption{Supervised UMAP embedding of the data, illustrating variance in both $X$ and $y$ (formation energy). The covariate shift data is located on the periphery, far from the majority of the data. The prior shift data forms its own distinct cluster, while the relation shift data is related but does not clearly fall into either of the previous categories.}
    \label{fig:data_distri}
\end{figure}

After removing all the OOD samples mentioned above, we constructed the conventional ML training and testing set using an 85:15 ratio to ensure that all test sets have roughly the same size. Finally, 10\% of the training set was used for validation. There are four main test sets for evaluation. All test sets were unseen during training. They are listed below, and their distribution is shown in Figure \ref{fig:data_distri}.
\begin{itemize}
  \item Independently and identically distributed (i.i.d.) test set
  \item Covariate shift test set
  \item Prior shift test set
  \item Relation shift (piezoelectric) test set
\end{itemize}

In addition, we also tested our models on special datasets with limited data size. We selected the vibration properties subset with 1,265 samples, the refractive index subset with 4,764 samples, and the shear stress subset with 10,987 samples for our experiments. Different guidelines were used to determine the OOD prior shift sets for these smaller datasets. Specifically, we selected the top 10\% and bottom 10\% of target values as our prior shift sets. Since these datasets are already subsets of materials with specific properties, there is no clear subset to select for relation shift, and thus, no relation shift sets were chosen.

\subsection{Crystal Adversarial Learning (CAL)}
Our CAL algorithm is adapted from the Stable Adversarial Learning (SAL) pipeline \cite{liu2021stable} and is integrated with individual residual networks for material property prediction. The core concept involves creating an uncertainty set by distinguishing features based on the stability of their correlations with the target. Adversarial samples are generated by perturbing unstable features using network gradients, and seed sample weights are optimized through iterative gradient descent.
Furthermore, partial sampling techniques are employed to iteratively select out-of-distribution (OOD) samples with high training losses. During this process, guided adversarial samples are generated, and partial sampling is used to enhance our model's ability to handle various OOD challenges.
\begin{figure}[h!]
\includegraphics[width=\textwidth]{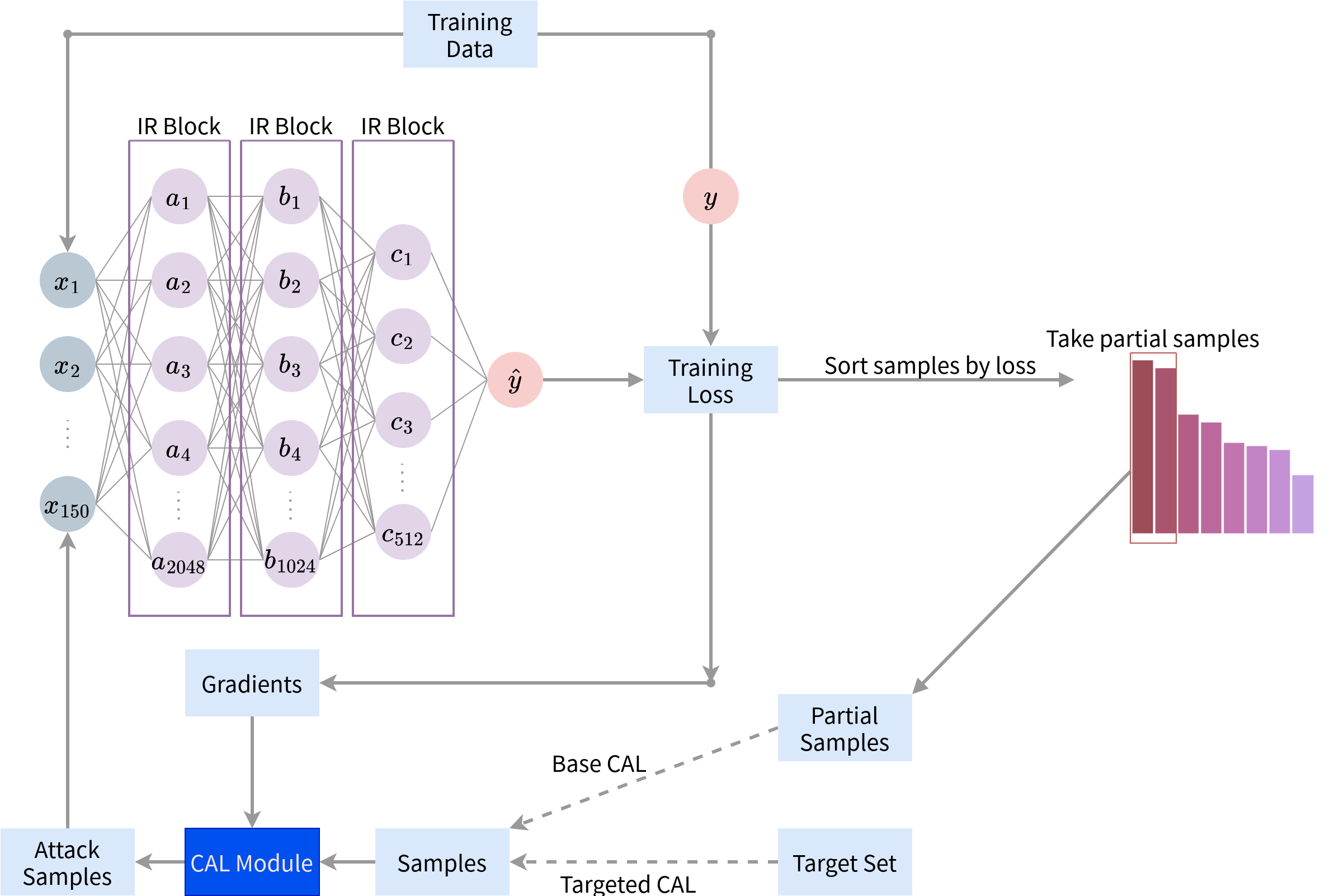}
\caption{The CAL pipeline for out-of-distribution (OOD) material property prediction. It is based on the Stable Adversarial Learning (SAL) pipeline and integrates individual residual networks for material property prediction. At each step, only one partial sample set and the corresponding target set are used for adversarial sample generation. The training schedule begins with five epochs of training on the entire dataset to initialize the network. Subsequently, the training samples are ranked based on their losses, and the top 20\% of high-loss samples are selected to continue training the network. In the base CAL model, these selected samples are used to generate adversarial attack samples. In our targeted training scheme, however, samples from the test domain are used to generate the adversarial attack samples.}
\label{fig:sal_pipeline}
\end{figure}

Figure \ref{fig:sal_pipeline} depicts the CAL pipeline and the experimental setup, where we first initialize the network with training samples for several epochs. Subsequently, the samples undergo sorting based on their loss at every $i^{\mathrm{th}}$ epoch, initiating at the $i^{\mathrm{th}}$ epoch. Concurrently, the CAL module extracts gradients as partial samples traverse the network. At every $j^{\mathrm{th}}$ epoch, the CAL module generates adversarial samples for the network, incorporating both generic and generated sample losses for adjustment. The training process continues until the average loss of i.i.d validation sets ceases to improve for $k$ consecutive epochs. This criterion is implemented to ensure the stability of the training distribution samples and marks the conclusion of the training phase. 
Another scenario is when working on a small data set, such as piezoelectric materials. The limited samples are a huge challenge for current ML prediction models.Fine-tuning with such a small dataset brings issues, such as overfitting and loss of robustness. Our CAL pipeline is able to target those datasets and generate adversarial samples based on them. In this setup, all adversarial samples are generated using one batch of piezoelectric samples, while others are not disrupted. In this way, the model has enough samples to learn from while avoiding overfitting.
We take an exponential moving average (EMA) of epoch checkpoints for our final model\cite{ema}. EMA is widely used for better generalization in normal training with SGD; we find that, for CAL training, EMA is even more useful. We hypothesize that CAL, with partial sampling rotating inputs periodically, induces training to go between different local minima representing different parts of the data. Averaging these weights at the end provides more robust performance than any individual model.

\paragraph{Targeting-Based Fine-Tuning for OOD SAL Algorithm} One of the major improvements in our CAL algorithm compared to SAL is the introduction of a targeting mechanism for fine-tuning a pretrained model to adapt to an OOD test set. In standard fine-tuning, a small batch of samples from the test set domain is typically used to fine-tune the model. However, this approach can easily lead to overfitting due to the limited number of fine-tuning samples. To address this, we propose a targeting-based fine-tuning mechanism, which uses selected test domain samples to guide the adversarial sample generation in SAL. This approach ensures that the generated adversarial samples are more similar to the test domain samples. The selected target domain samples serve as the seed samples for adversarial sample generation, in contrast to the original SAL, which uses all training samples as seeds. This modification not only makes our CAL algorithm more scalable but also guides the weighting towards target domain samples, significantly improving OOD performance compared to standard fine-tuning.

\paragraph{Partial Sampling} We adapted the partial sampling approach from Magar \textit{et al.} \cite{magar2023learning}. Periodically during training, the training set is sorted based on the loss computed by the current model. The batches with the highest loss, referred to as partial samples, are the only training batches used until new partial samples are regenerated. This entire process is depicted in Figure \ref{fig:sal_pipeline}.

\paragraph{Layer Block in Our CAL Neural Network Model}
We adopted and modified IR-net \cite{jha2019irnet} as the backbone for our model, as shown in Figure \ref{fig:sal_pipeline}. The original IR-net achieved state-of-the-art results with an intricate architecture comprising 48 layers of individual residual blocks. However, for our data, we found that this depth was counterproductive, so we opted for a more compact architecture.

\begin{figure}[t!]
   \includegraphics[width=\textwidth]{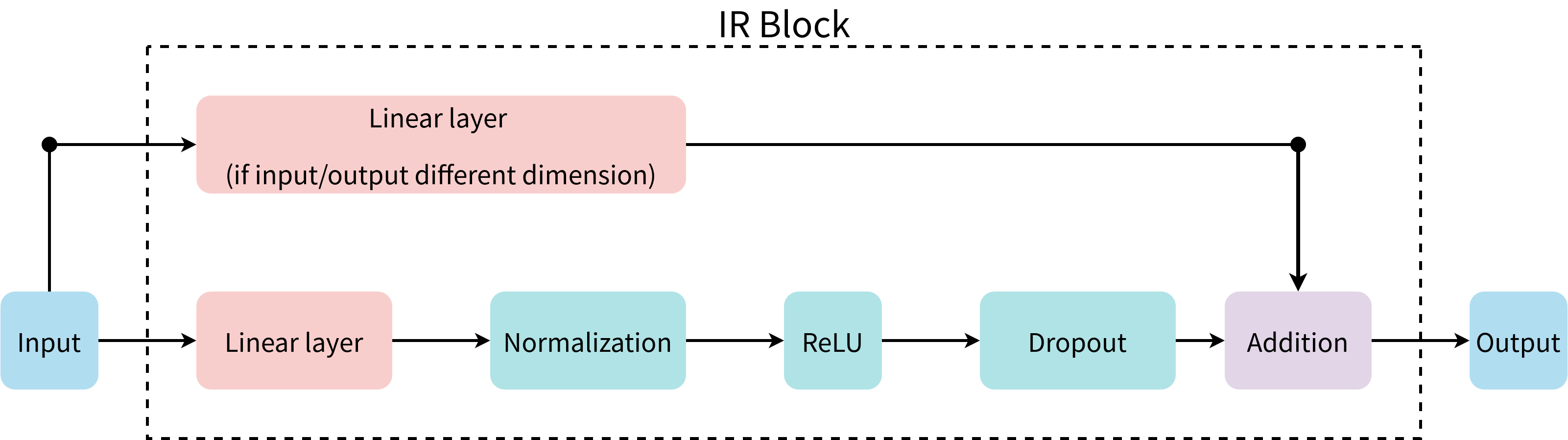}
    \caption{Neural network architecture of the IR-net block. The primary difference from a standard multi-layer perceptron is the addition of the outputs to the inputs. If the input and output dimensions differ, a linear projection of the inputs is used instead.}
    \label{fig:IRblock}
\end{figure}

In our refinement, we streamlined the network to three layers with individual residual blocks, having dimensions of 2048, 1024, and 512, respectively. The structure of the individual residual blocks is shown in Figure \ref{fig:IRblock}. Unlike conventional residual neural networks, we add the original input to the output after each block. The input vector is first processed through a fully connected layer, followed by layer normalization (we use LayerNorm \cite{layernorm} instead of the original batch normalization) and ReLU activation. Additionally, we found that dropout is essential for mitigating overfitting in our dataset. Our experiments demonstrate that our improved IR-net, which is wider but shallower, achieves similar or even better prediction performance compared to the original IR-net, with improved computational efficiency.

\FloatBarrier

\paragraph{Adversarial Attack Algorithm for OOD Machine Learning (AA)}\label{sec:M_sal}
We assimilated the SAL algorithm from Liu \textit{et al.} \cite{liu2021stable} for OOD materials property prediction. The SAL algorithm constructs an uncertainty set by categorizing covariates into stable and unstable groups based on their correlation with the target variable. It generates adversarial samples by perturbing the unstable covariates (features) using network gradients and optimizes covariate weights through iterative gradient descent to balance both average and worst-case performance across training environments. It is important to note that the algorithm relies on the availability of the training domain distribution to identify covariate feature vectors. The original algorithm also utilized Wasserstein Distributionally Robust Learning (WDRL), which aims to create an uncertainty set characterized by the stability of covariates, thereby inducing stronger adversarial perturbations on unstable covariates. In their work, the algorithm was tested on a real-world house price prediction problem using simple, fully connected single-layer perceptrons. However, certain challenges remain unresolved. Due to the nature of this method, the original adversarial data generation was not optimized for computational complexity. As a result, it has limited scalability, and its performance on large, complex datasets remains unknown.

From the framework of \textit{Wasserstein Distributionally Robust Learning} (WDRL) \cite{esfahani2015data}, the original SAL architecture further characterizes the uncertainty set as anisotropic based on the stability of covariates across multiple environments. This induces stronger adversarial perturbations on unstable covariates than on stable ones. The WDRL framework is built on two key assumptions:
\textbf{Assumption 1} suggests that there exists a decomposition of all the covariates, \emph{X = \{S, V\}}, where $S$ represents the stable covariate set and $V$ represents the unstable one. The correlation between stable covariates $S$ and the target $Y$ remains invariant across environments \cite{arjovsky2019invariant}. This is supported by the success of deep learning models in material property prediction, where high accuracy rates indicate the presence of stable covariates that the network can learn.
\textbf{Assumption 2} posits that there exists at least one distribution on the boundary whose marginal distribution on $S$ differs from the center distribution of the original uncertainty set \cite{liu2021stable}. This assumption is satisfied by the artificially created distribution shift in our OOD problem.

If stable covariates are perturbed in an adversarial learning process, the target ($y$) should change proportionally to preserve the underlying relationship. In practice, however, we do not have access to the target values corresponding to the perturbed stable covariates. Consequently, optimizing under an isotropic uncertainty set that includes perturbations on both stable and unstable covariates typically reduces the model's confidence and produces meaningless results. This issue is addressed by assigning higher weights to stable covariates in the cost function, thereby constructing a more reasonable uncertainty set that avoids ineffective perturbations. The combined objective function below optimizes both the covariate differentiation process and the adversarial training of model parameters:
\begin{equation}
\min _{\theta \in \Theta} \sup _{Q: W_{c_w}\left(Qcharts, P_0\right) \leq \rho} \mathbb{E}_{X, Y \sim Q}[\ell(\theta ; X, Y)]
\end{equation}
where\begin{equation}
c_w\left(z_1, z_2\right)=\left\|w \odot\left(z_1-z_2\right)\right\|_2^2 
\end{equation}and
\begin{equation}
w \in \arg \min _{w \in \mathcal{W}}\left\{\frac{1}{\left|\mathcal{E}_{t r}\right|} \sum_{e \in \mathcal{E}_{t r}} \mathcal{L}^e(\theta)+\alpha \max _{e_p, e_q \in \mathcal{E}_{t r}} \mathcal{L}^{e_p}-\mathcal{L}^{e_q}\right\}
\end{equation}
\(P_{0}\) denotes the training distribution, 
\(W_{c_{w}}\) denotes the Wasserstein distance with transportation cost function
\(c_{w}\) defined as equation 3, 
\(\mathcal{W}=\) \(\left\{w: w \in[1,+\infty)^{m+1} \& \& \min \left(w^{(1)}, \ldots, w^{(m+1)}\right)=1\right\}\) denotes the covariate weight 
space(\(w^{(i)}\) denotes the $i^{\mathrm{th}}$ element of \(w\)), 
and \(\mathcal{L}^{e}\) denotes the average loss in environment 
\(e \in \mathcal{E}_{t r}\). $\alpha$ is a hyperparameter that trades off between average performance and the stability\cite{liu2021stable}.

The adversarial sampling objective function is:

\begin{equation}
R(\theta(w))=\frac{1}{\left|\mathcal{E}_{t r}\right|} \sum_{e \in \mathcal{E}_{t r}} \mathcal{L}^{e}(\theta(w))+\alpha \max _{e_{p}, e_{q} \in \mathcal{E}_{t r}}\left(\mathcal{L}^{e_{p}}-\mathcal{L}^{e_{q}}\right)
\end{equation}

where \(\alpha\) is the hyper-parameter. 
\(R(\theta(w))\) contains two parts:
the first is the average loss in multiple training environments; 
the second reflects the max margin among environments, which reflects the stability of \(\theta(w)\), 
since it is easy to prove that \(\max _{e_{p}, e_{q} \in \mathcal{E}_{t r}} \mathcal{L}^{e_{p}}(\theta(w))-\mathcal{L}^{e_{q}}(\theta(w))=0\) if and only if the errors among all training environments are same. 
Originally \(\alpha\) is used to adjust the trade off between average performance and stability.
\paragraph{Technical improvements for increasing SAL scalability:}
The original SAL algorithm \cite{liu2021stable} was designed for small models and datasets, allowing for the straightforward calculation of all necessary gradients. However, our datasets and models are significantly larger, necessitating several modifications to make SAL feasible. Instead of generating an adversarial copy of the entire dataset, we sub-sample a portion for adversarial data generation. This approach is analogous to the standard use of stochastic gradient descent (SGD) rather than full gradient descent. Additionally, to avoid the memory-intensive task of materializing the full Jacobian, we compute only the necessary Jacobian-vector products.

The gradient used to optimize $w$, the SAL perturbation weights on features in the dataset, is given as equation 11 in \cite{liu2021stable}, where $X_A$ is the adversarial data and $\theta$ represents the model parameters:
\begin{equation} \label{sal-loss}
    \frac{\partial R(\theta(w))}{\partial w} = \frac{\partial R}{\partial \theta} \frac{\partial \theta}{\partial X_A} \frac{\partial X_A}{\partial w}
\end{equation}
The original SAL implementation stores $\frac{\partial \theta}{\partial X_A}$ in memory before computing this product. For models with large parameter counts, this full Jacobian does not fit in memory as it contains $\dim(\theta) \times \dim(x)$ entries. Our implementation addresses this limitation by computing the vector-Jacobian product $\frac{\partial R}{\partial \theta} \frac{\partial \theta}{\partial X_A}$ directly, which has only $\dim(x)$ entries. This optimization enables the application of SAL to models with many thousands of parameters without encountering memory issues.

The loss in (\ref{sal-loss}) can be further approximated by using only a subset of the parameters $\theta$. Although such approximations are not necessary for our work after the optimizations discussed above, we explore their effects in our ablation study by approximating the gradient using the linear weights of the first or last layer. Additionally, we find that normalizing the input $X$ can enhance the scalability of SAL. All these technical improvements have enabled us to deploy SAL effectively in modern deep learning networks for material property prediction, as utilized in our CAL model.

In this study, we implemented smoothed $L_1$ loss, also known as Huber loss, due to its advantageous properties in machine learning applications. This loss function effectively combines the strengths of $L_1$ and $L_2$ losses, providing robustness against outliers while simultaneously maintaining gradient stability. For small errors, it behaves like $L_2$ loss, ensuring smoothness and differentiability near zero. For larger errors, it transitions to $L_1$ loss characteristics, thereby mitigating the impact of outliers. This dual nature not only prevents gradient explosion during training but also often results in accelerated and more stable convergence compared to pure $L_1$ loss. Given these attributes, smoothed $L_1$ loss is particularly effective for regression tasks, making it an optimal choice for our CAL pipeline training. Simultaneously, we used standard mean absolute error (MAE) for test set evaluation.

\section{Results}\label{sec:Results}
\FloatBarrier
\subsection{Standard ML models have poor performance on OOD datasets}
\begin{table}[htb!]
  \begin{center}
    \caption{Performance of state-of-the-art models on our out-of-distribution (OOD) test sets for formation energy prediction (units: eV), measured in mean absolute error (MAE).}
    \label{tab:SoArt_OOD}
    \begin{threeparttable}[b]
      \begin{tabular}{l|c|c|c}
        \toprule
        \textbf{Data Set}\tnote{1} & \textbf{CrabNet\tnote{2}} & \textbf{MODNet} & \textbf{IRNet-48} \\
        \midrule
        \textbf{Training Set}                & 0.0412 & 0.066 & 0.002 \\
        \textbf{Test Set (i.i.d)}            & 0.0624 & 0.125 & 0.411 \\
        \textbf{Covariate Shift}             & 0.260  & 0.2659 & 0.408 \\
        \textbf{Prior Shift}                 & 0.953  & 0.9089 & 2.091 \\
        \textbf{Relation Shift}              & 0.571  & DNF\tnote{3} & 1.188 \\
        \bottomrule
      \end{tabular}
      \begin{tablenotes}
        \item [1] The dataset $x$ attribute is not normalized in this table.
        \item [2] CrabNet generates all features internally from the formula.
        \item [3] MODNet performs feature selection, and the Piezoelectric materials are missing some features as NaN. The feature selector removes columns that cause misalignment between the dataset dimension and the network dimension.
      \end{tablenotes}
    \end{threeparttable}
  \end{center}
\end{table}
Table \ref{tab:SoArt_OOD} summarizes the performance of three deep learning models—CrabNet, MODNet, and IRNet-48 \cite{wang2021compositionally,de2021materials,jha2019irnet}—across various test sets, highlighting their susceptibility to out-of-distribution (OOD) data. All models achieve low error rates (around 0.002–0.06 eV) when trained and validated on the training dataset ("Training Set") and the i.i.d test set. However, their performance significantly degrades on our OOD datasets ("Covariate Shift", "Prior Shift", "Relation Shift"), with error rates increasing substantially across all models, ranging from 0.224 eV to as high as 2.091 eV.

The models' performances are acceptable over the random split and covariate shift datasets. The covariate shift does not pose a significant challenge for neural network models, as the shift in the $X$ domain involves physical real-world data where the OOD samples are not far from the majority of the data distribution. In contrast, the prior shift selects OOD samples in the $Y$ domain that are unfamiliar to the model, although their $X$ features are still within the training distribution, as shown in Figure \ref{fig:data_distri}. The blue dots in the figure represent the prior shift samples, which overlap with the training samples shown as grey dots. The relation shift, marked in yellow in Figure \ref{fig:data_distri}, is characterized by limited samples. The piezoelectric materials selected for this shift are considered special within materials science studies, even though their $X$ and $Y$ are in-distribution. This is likely why the relation shift performance falls between that of the prior shift and covariate shift for the benchmark models.

Model-wise comparisons reveal that CrabNet and MODNet outperform the traditional neural network model IRNet-48. CrabNet, which uses attention-based element and atomic number representations, and MODNet, which employs hierarchical properties grouping, exhibit better performance than IRNet-48. IRNet-48 demonstrates a severe overfitting problem, with a training loss as low as 0.002 eV compared to a validation loss (random splitting) of 0.441 eV. This indicates that all three models struggle with generalizing to unseen data distributions, underscoring that our OOD dataset presents a meaningful challenge for current models.

\subsection{OOD Performance of the CAL algorithm}
\begin{table}[h!]
  \begin{center}
    \caption{
    Out-of-Distribution (OOD) Performance (MAE) Comparison: Our CAL method maintains stability during training compared to the baseline for predicting formation energy (units: eV). For all targeted datasets, our CAL pipeline consistently outperforms traditional fine-tuning methods across various OOD shifts. Training was performed using smoothed L1 loss, and model performance was evaluated using mean absolute error (MAE).
    }
    \begin{threeparttable}[b]
    \begin{tabular}{r|r r r r r}
       \toprule %
                    & Train  & Random   & Covariate         & Prior     & Relation  \\
                    &  Set    &  Split   & Shift             & Shift     & Shift \\
      \midrule %
        IR-net Baseline\tnote{1}   & 0.0009       & 0.0863      & 0.2448     & 1.3672    & 0.2222   \\
        Our CAL Baseline           & 0.2111       & 0.5424      & 0.4925     & 1.3298    & 0.6115   \\
        \midrule %
        IR-net Fine-tune \tnote{2}  & -             & -             & 0.2005     & 0.1471    & 0.1907    \\
        Targeted CAL \tnote{2}      & -             & -             & 0.1701     & 0.1216    & 0.0932    \\
        \midrule
        Fine-tune \% Improvement    & -             & -             & 18.09\%    & 89.24\%   & 14.17\%   \\
        Targeted CAL \% Improvement & -             & -             & 65.46\%    & 90.85\%   & 84.75\%   \\
       \bottomrule %
    \end{tabular}
        \begin{tablenotes}
           \item [1] This model serves as a baseline for fine-tuning.  
           \item [2] Each column in these rows corresponds to different fine-tune/target sets.
        \end{tablenotes}
    \end{threeparttable}
    \label{tab:GeneralPerformanceComparison}
  \end{center}
\end{table}

Table \ref{tab:GeneralPerformanceComparison} summarizes the OOD performance of the baseline models and fine-tuning models, comparing our CAL method with the conventional IR-net model. When comparing the IR-net baseline with CAL on the base dataset using i.i.d. random splits, we observe that the IR-net baseline performs well on both the random splits and the covariate shift datasets. This is consistent with previous observations that covariate shifts are not particularly challenging for neural network models, as shown in Table \ref{tab:SoArt_OOD}. This performance indicates that neural networks tend to overfit to the training distribution. Consequently, the performance on OOD test sets is significantly lower than on i.i.d. datasets, highlighting the overfitting issue. In contrast, the CAL model demonstrates more consistent performance across OOD scenarios. While it shows a higher mean absolute error (MAE) of 0.5424 eV on the random split set, 0.4925 eV on the covariate shift set, and 0.6115 eV on the relation shift set, it maintains better performance stability across OOD data compared to the IR-net baseline. This stability reflects the effectiveness of the CAL approach in handling OOD data, even at the cost of slightly reduced performance on data similar to the training dataset.
It is interesting to note that our CAL baseline performs best on the covariate shift set, even surpassing the performance on the i.i.d. random split set. This can be attributed to the combination of partial training and adversarial attacks. The covariate shift samples are selected by identifying outliers in the feature $X$ domain. Although these samples are removed and unseen by the CAL model, they have neighbors in the feature map that remain in the training set. These samples exhibit high losses when processed through the network due to their deviation from the majority of the sample distribution. The partial sampling mechanism effectively selects these outliers as seed samples for adversarial attacks. Compared to the IR-net baseline, which shows clear overfitting under the same training hyperparameters, our CAL baseline demonstrates improved performance on the prior shift set. This performance on covariate shift indicates the substantial potential of CAL for handling OOD scenarios.

Our CAL method facilitates adversarial learning using a specific secondary training set rather than sampling from the original training set. This approach requires a base training set and a secondary, smaller set of \textit{labeled} training data from a specific domain where we aim to improve model performance. The standard approach for this problem in the literature is fine-tuning: training a base model and then updating it with a new training set. We investigate whether our targeted CAL can outperform standard fine-tuning when we have a small number of labeled examples from the target distribution, in addition to a larger training set with potential distribution shifts. Table \ref{tab:GeneralPerformanceComparison} shows the performance of these models at the bottom. For larger OOD covariate shift and prior shift datasets (with 1024 samples used for training), our model achieves modest performance improvements over the baseline, with 65.46\% and 90.85\% lower losses, respectively. 
Meanwhile, the targeted CAL outperforms the IR-net fine-tune by 21.3\% on the prior shift test set, with a higher improvement ratio. Considering that our targeted CAL, unlike fine-tuning, does not require a baseline model to start training, achieving similar or even better results demonstrates the capability of our CAL model.
For the piezoelectric data, with only 256 samples for training, we observe a much more significant improvement: a 51.12\% lower loss for the targeted CAL, with an 84.75\% improvement over the baseline, whereas the fine-tune method only improved by 14.17\% over the baseline.

Additionally, we gained insights into our OOD dataset by evaluating the performance of targeted versus fine-tuned results. Figure \ref{fig:data_shift} (b) indicates that the majority of prior shift samples are near the training set in terms of $Y$ values, and Figure \ref{fig:data_distri} shows the overlap between prior shift and training set samples. We believe this explains why the IR-net fine-tune model performs better on the prior shift than on the covariate shift. In contrast, the targeted CAL generates adversarial samples that guide the model towards better stability. It is also worth noting that with small datasets, the fine-tuning strategy can lead to severe overfitting, resulting in performance degradation compared to the baseline, as indicated by the relation shift test results in Table \ref{tab:SoArt_OOD}.

\begin{table}[htb!]
  \begin{center}
    \caption{
    The targeted CAL model demonstrates better MAE OOD performance in bandgap prediction (eV) across all OOD shift test sets.}
    \begin{threeparttable}[b]
    \begin{tabular}{r|r r r r r}
       \toprule %
                    & Train  & Random   & Covariate & Prior   & Relation \\
                    &        & Splits   & Shift     & Shift   & Shift \\
      \midrule %
        IR-net Baseline     & 0.0259 & 0.3214   & 0.6178    & 2.8085  & 0.8176 \\
        CAL Baseline        & 0.4841 & 0.9242   & 1.0686    & 3.0002  & 1.0649 \\
        \midrule %
        IR-net Fine-tune    & -      & -        & 0.4705    & 1.0276  & 0.7602 \\
        Targeted CAL        & -      & -        & \textbf{0.3521} & \textbf{0.4425} & \textbf{0.2524} \\
        \midrule
        \% Fine-tune Improvement & -    & -        & 34.04\%   & 63.41\% & 7.02\% \\
        \% CAL Improvement      & -    & -        & 67.05\%   & 85.25\% & 76.29\% \\
       \bottomrule %
    \end{tabular}
    \end{threeparttable}
  \label{tab:bandgap}
  \end{center}
\end{table}

We also evaluated the performance of targeted CAL for bandgap prediction and compared it with the baseline CAL algorithm without targeting. The OOD test sets were selected from the 84,190 MP samples, as discussed in the Dataset section. In this experiment, our prior shift set was selected using bandgap energy, and the performance results are shown in Table \ref{tab:bandgap}. We observe similar and even better performance advantages compared to the fine-tuning method, as shown previously. Not only does our targeted CAL outperform the fine-tuning method on all OOD shift sets, but it also shows better improvement over its baseline model. In terms of data distribution, we see a significant performance increase for both models on the prior shift set, with more than 60\% improvement over the baseline for both models. As before, the fine-tuning method shows a minor improvement on the relation shift test set, with a 7.02\% improvement rate, indicating overfitting on the visible dataset. In contrast, our targeted CAL achieved a 66.79\% better performance. This dataset, demonstrating the largest improvement of our method for both bandgap and formation energy prediction and informed directly by existing crystal OOD challenges, merits further consideration.

\FloatBarrier
\subsection{Piezoelectric materials: a case study}
\begin{table}[h!]
    \begin{center}
    \caption{Piezoelectric prediction (MAE) performance comparison: Smoothed mean squared error for models predicting formation energy (eV) on piezoelectric data, comparing targeted CAL with fine-tuning.}
    \begin{tabular}{r | r r r r }
    \toprule
      \textbf{Method} & \textbf{CAL} & \textbf{Fine-tune} & \textbf{CAL} & \textbf{Fine-tune} \\
      \textbf{\# Finetuning Samples} & 128 & 128 & 256 & 256 \\ 
      \midrule
      \textbf{Loss (Held-out Piezoelectric)} & 0.0774 & 1.4488 & \textbf{0.0443} & 0.0877 \\
    \bottomrule
    \end{tabular}
    \label{tab:PiezobatchPerformance}
    \end{center}
\end{table}

Due to the small size of the dataset (317 samples), we investigate whether varying the amount of samples held out for validation affects the results. Table \ref{tab:PiezobatchPerformance} shows that fine-tuning fails to leverage additional data effectively, whereas our modified CAL continues to improve with an extra 128 training data points used to generate adversarial samples. 
Given the small dataset, we hypothesize that CAL functions as a form of data augmentation, providing the model with more data to learn from. This newly generated adversarial data helps prevent overfitting and enhances learning. Future work could explore other data augmentation techniques for low-data regimes in materials science and evaluate our method's performance on additional small materials science datasets.
Ultimately, any model will perform optimally on a specific subset of data, underscoring the importance of precise assessment when applying machine learning models for downstream material property prediction tasks. Training and validation set performance in these experiments do not strongly correlate with performance on specific types of downstream test data with distribution shift.

\FloatBarrier
\subsection{CAL Performance over Datasets with Limited Samples}
To further test the capability of our CAL model, we selected datasets from Matbench \cite{dunn2020benchmarking} with limited samples available for different properties, such as phonon vibration properties, shear modulus, and refractive index. Here, we focus on CAL performance in predicting phonon or vibration properties, given their limited sizes and challenging target value distribution with a median absolute deviation of 323.78. Results for other datasets are available in the supplementary files, Tables S1 and S2. We tested the performance on the vibration properties dataset with 1,265 samples, the refractive index dataset with 4,764 samples, and the shear stress dataset with 10,987 samples. We used different guidelines to determine the OOD test sets for these small datasets. We selected samples with the top 10\% and bottom 10\% of target values as our prior shift set. Since these datasets are already subsets of materials with specific properties, there is no clear subset to select for relation shifts, and thus no relation shift sets were chosen. 

\begin{figure}[h!]
    \centering
    \begin{subfigure}[b]{0.49\textwidth}
        \centering
        \includegraphics[width=1\linewidth]{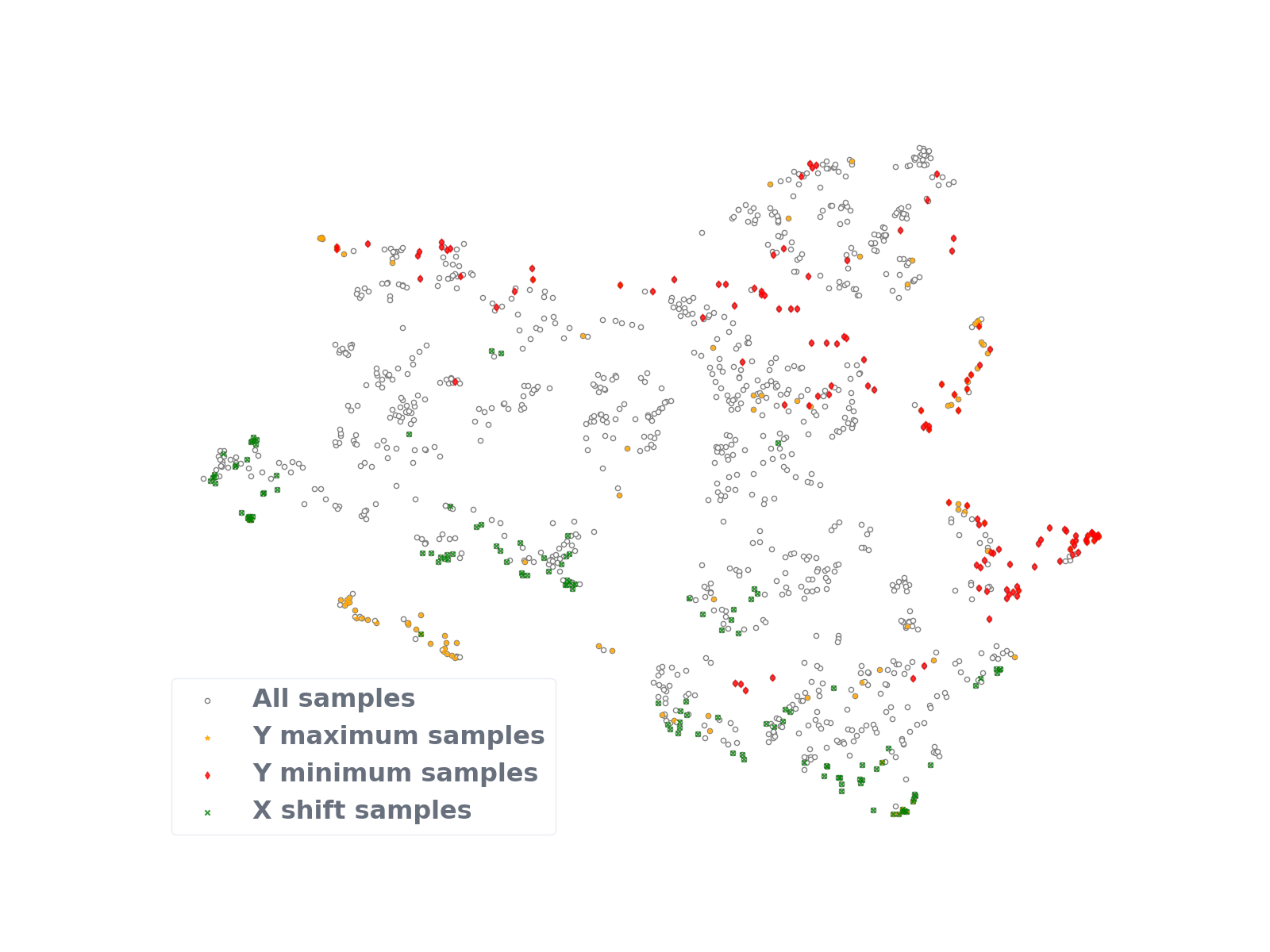}
        \caption{2D feature map visualization of OOD shift samples}
    \end{subfigure}
    \begin{subfigure}[b]{0.49\textwidth}
        \centering
        \includegraphics[width=1\linewidth]{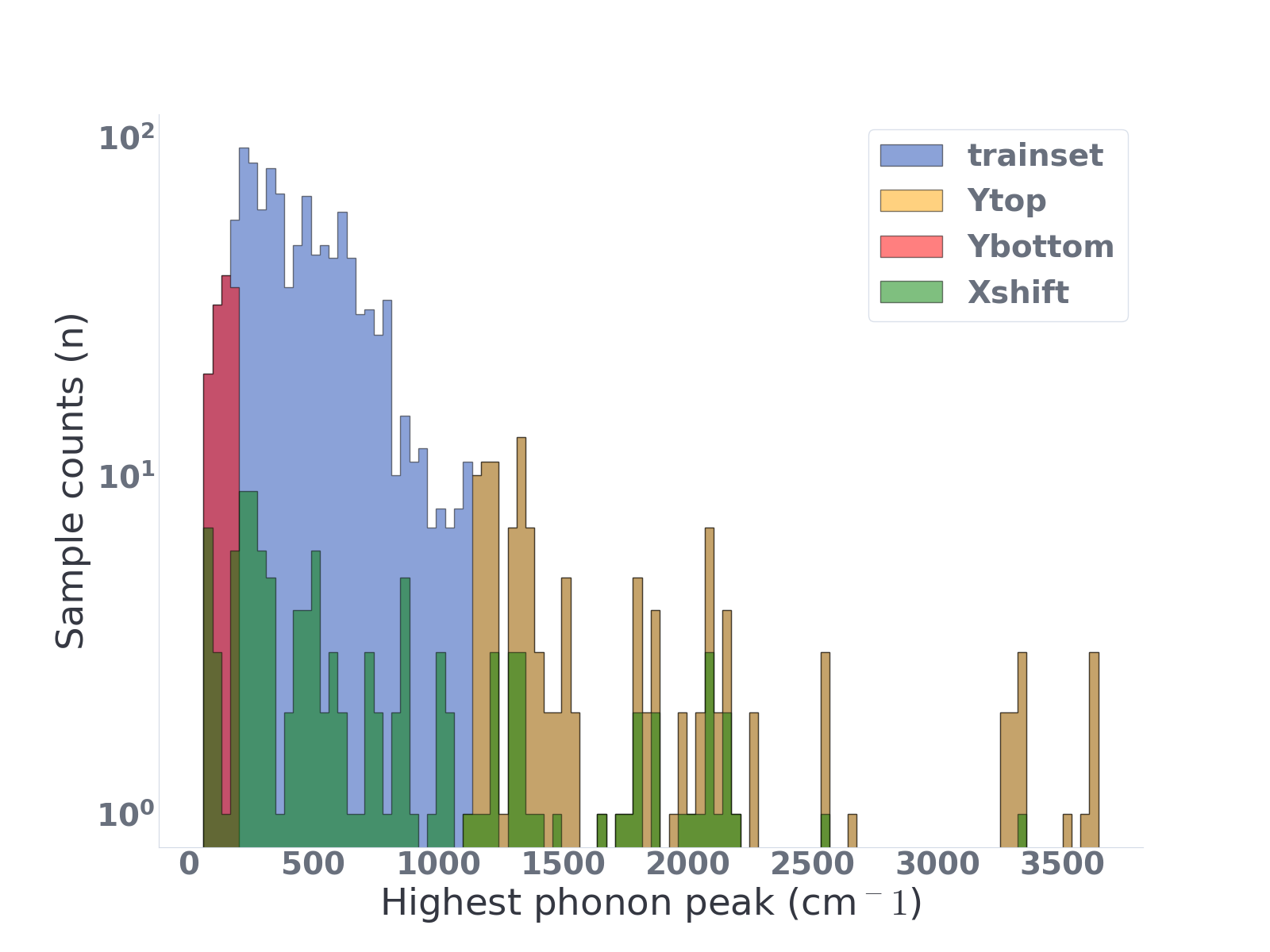}
        \caption{OOD test sets target value distribution}
    \end{subfigure}
    \caption{Visualization of OOD test set samples: (a) Two-dimensional compositional feature space representation, with colored markers indicating individual samples' relationships to local clusters and global structure. (b) Distribution of target values for the OOD test sets, showing the frequency of the highest optical phonon mode peak. The y-axis is on a log scale.}
    \label{fig:phonon_shift}
\end{figure}

Figure \ref{fig:phonon_shift} (a) shows a two-dimensional projection of the data using UMAP, with test sets distinguished by colors. This projection clearly separates the test sets, indicating distribution shifts. Notably, the values with low $y$ are more heterogeneous, suggesting that low peaks may not be as indicative of composition as high peaks. Figure \ref{fig:phonon_shift} (b) illustrates the frequency of the highest frequency optical phonon mode peak, in units of $cm^{-1}$, for this vibration properties prediction subset. The majority of samples have their phonon peak distributed between 0 $cm^{-1}$ and 1,500 $cm^{-1}$, while some samples have peaks as high as 3,643.74 $cm^{-1}$. With a median absolute deviation of 323.78, even the best model on the Matbench leaderboard (as of May 2024) only achieved a mean absolute error of 28.76 under standard benchmark conditions. Our method of identifying the OOD shift set adds to the challenge. For this experimental setup, our training set contains 799 samples, our regular test set includes 130 samples, and each OOD shift test set comprises 126 samples.

As shown in Table \ref{tab:phonon}, the IR-net baseline achieves a smooth L1 training loss of 13.77 and a testing loss of 43.71. Compared to the benchmark leaderboard, the test loss falls into the 10th place (as of May 2024), which is reasonable. On the other hand, the CAL baseline struggles, with a high training loss of 142.14 and a test loss of 143.63. It is just behind the random forest on the leaderboard. We suspect this is due to the nature of our CAL pipeline and the distribution of the data. As shown in Figure \ref{fig:phonon_shift} (b), most minimum prior shift data are fairly near the training set distribution, hence the best loss of 49.62 in all test sets for the CAL baseline model. We also observed earlier that there is an overlap of samples from two prior shift sets in Figure \ref{fig:phonon_shift} (a). This suggests a complex relationship between the selected composition features and phonon peaks. This means that the hard-to-learn samples in the training set may not be similar to the OOD test sets. At the same time, the limited dataset siz
Due to the limited size of the test sets, we opted for a batch size of 32 samples and used two batches for fine-tuning or targeting. Our CAL Fine-tune model achieves significantly better results compared to the IR-net Fine-tune model across all three OOD test sets, even though its baseline performs poorly. For the X shift set, standard neural network machine learning models tend to have good performance, as shown previously. For phonon vibration prediction, however, the IR-net fine-tune model struggles to improve, ending up only 5.55\% better than the baseline, while our CAL fine-tune has a loss of 76.10, becoming 82.91\% better than the CAL baseline. For the maximum prior shift set, both baseline models have comparably high losses. Our targeted CAL model not only outperforms the IR-net fine-tuning but also achieves an i.i.d. test set loss of 62.85 after fine-tuning, compared with the IR-net fine-tune model's i.i.d. test loss of 134.52. This demonstrates the robustness of our model after fine-tuning. Finally, our model also outperforms the IR-net's strong baseline for the minimum prior shift loss, with a 45.32\% improvement and a loss of 27.13. All results suggest that although unrepresentative samples may cause significant performance drops, the partial training and adversarial attack module in our model are highly effective in problems with limited sample sizes when used to target a specific subset of data, akin to fine-tuning.
\begin{table}[h!]
  \begin{center}
    \caption{OOD (MAE) performance comparison: models predicting vibration properties last phonon peak ($cm^{-1}$). 
            Smoothed mean squared error is used during training and reported for the train set. The test set and all OOD datasets are evaluated by mean MAE. 
            This dataset is challenging due to the limited samples (1265) and extreme target value distribution with a median absolute deviation of 323.78. 
            Our CAL Fine-tune model achieves significantly better results compared to the IR-net Fine-tune model across all three OOD test sets even though its baseline performs poorly.}
    
    \begin{threeparttable}[b]
        \begin{tabular}{r|r r r r r}
        \toprule %
                     & Train   & Test    & Covariate     & Maximum prior & Minimum prior \\
                     & set     & set    & shift        & shift        & shift \\
        \midrule %
        IR-net baseline      & 13.77 & 43.71        & 232.90 & 770.39 & 38.42     \\
        Our CAL base         & 142.14 & 143.63      & 445.53 & 1231.59 & 49.62  \\
        \midrule %
        IR-net Fine-tune   & -     & -     & 219.96        & 523.35           & 23.47 \\
        Our Targeted CAL   & -     & -     & \textbf{76.10} & \textbf{195.23} & \textbf{27.13}    \\
        \midrule
        Fine-tune \% improvement             & -   & -   & 5.55\% & 32.06\% & 38.91\% \\
        Targeted CAL \% improvement          & -   & -   & \textbf{82.91\%} & \textbf{84.14\%} & \textbf{45.32\%} \\
        \bottomrule %
        \end{tabular}
    \end{threeparttable}
  \label{tab:phonon}
  \end{center}
\end{table}
\FloatBarrier

\subsection{Ablation Study}\label{sec:ablation}
\paragraph{Effects of the Partial Sampling and Adversarial Attack}
To examine the effect of the proposed CAL pipeline, we compare the CAL, just Adversarial Attack (AA), just Partial Sampling (PS), and the baseline neural network (IRnet-3) as shown in Table \ref{tab:ablation_study}. Compared to the baseline "IRnet-3" model, our CAL consistently achieves similar or lower loss values across all data conditions (Training Set, Random Splitting, Covariate Shift, Prior Shift, Relation Shift). This suggests that the combination of Adversarial Attacks (AA) and Partial Sampling (PS) within our approach effectively mitigates the negative effects of data distribution shifts. As mentioned earlier, neural network models have certain stability against the covariate shift, while a simple adversarial attack brings more loss to the model if not well guided, hence the increase in loss. On the other hand, the Partial Sampling module alone shows some effectiveness against the covariate shift set. This is intuitively true since it is pushing the model to focus on high-loss samples where the $Y$ values are in distribution and their $X$ values may not be, adding an element of aleatory uncertainty. At the same time, the distribution of the formation energy in our prior set (Figure \ref{fig:data_shift} (b)) indicates that the majority of prior shift samples are near the training set in terms of $Y$ value. The Partial Sampling technique can be beneficial for improving the performance of edge data from the training set, hence the leading result of the PS model for the Prior shift set. Finally, we notice that the AA module is highly efficient when dealing with extremely small datasets. The generated adversarial samples can be highly relevant to the target samples. The lack of data becomes advantageous. This finding prompted us to further improve our CAL pipeline by introducing the targeted training procedure.

\begin{table}[h!]
    \centering
    \caption{Smoothed MAE performance of our baseline CAL model and its ablated versions. AA stands for the model with only adversarial attacks and PS for the model with only partial sampling. IRnet-3 is the barebone neural network model. We find partial sampling alone is less effective. Adversarial attack is effective when dealing with extremely small datasets.}
    \begin{tabular}{c|rrrrr}
        \toprule %
        Method     & Training Set           & Random Splitting          & Covariate Shift          & Prior Shift           & Relation Shift \\ 
        \midrule %
        
        Our CAL    & 0.0061                 & 0.0851                    & 0.0856                   & 0.1564               & 0.0781                   \\
        AA         & 0.0163                 & 0.0915                    & 0.1067                   & 0.1522               & \textbf{0.0773}                  \\
        PS         & 0.0081                 & 0.0853                    & 0.0900                   & \textbf{0.1403}               & 0.0801                  \\
        IRnet-3    & 0.0059                 & 0.0838                    & \textbf{0.0837}                   & 0.1676               & 0.0823                 \\
        \bottomrule %
    \end{tabular}
    \label{tab:ablation_study}
\end{table}

\begin{figure}[h!]
    \centering
    \includegraphics[width=0.8\textwidth]{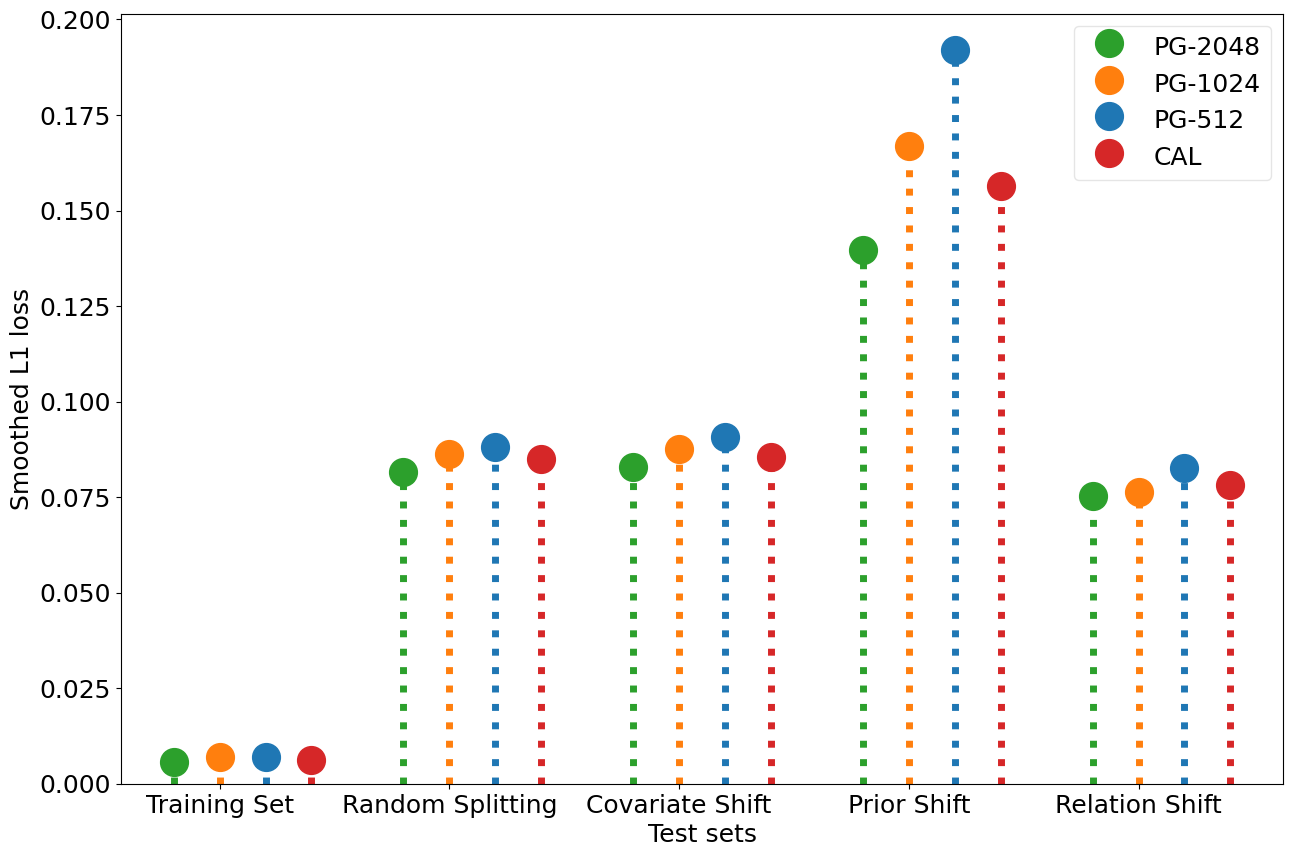}
    \caption{Comparison of smoothed Mean Absolute Error (MAE) performance for our baseline CAL model using partial gradients. The backbone neural network of our CAL consists of three layers with a 2048-1024-512 structure, and each color in the graph represents a partial gradient experiment focused on a specific layer. For instance, PG-2048 denotes the experiment where gradients for adversarial attacks are exclusively extracted from the 2048-unit layer.}
    \label{fig:abl_grad}
\end{figure}

\paragraph{One Layer Adversarial Gradient Attack}
At the early stage of our experiment, we encountered a computational cost problem, where the gradient size increased explosively with the size of the dataset and the complexity of the network. This is also the reason why we did not focus on more popular and complex structure-based crystal property prediction problems. To address the computational cost issue of our CAL model for complex neural network models, we proposed and evaluated the OOD performance for adjusted CAL models where gradients are derived from only one layer of the network.
The Figure \ref{fig:abl_grad} shows the performance comparison of CAL models using gradients from different layers of the adversarial attack module, compared with the full-gradient CAL model. The backbone neural network of our CAL consists of three layers with a 2048-1024-512 structure. PG-2048 refers to the experiment where gradients for the adversarial attack are exclusively extracted from the 2048-unit layer, which demonstrates the best OOD performance among the tested configurations. In the in-distribution random splitting test set, we observed a gradual increase in loss as the gradient calculation moved from layer 2048 to layer 512, with our CAL baseline performance falling between the 2048 and 1024 layers. The gradient calculation from the first layer, which corresponds to feature extraction \cite{li2023global}, yields the best performance among all layers. This pattern is consistent across the covariate shift and prior shift test sets. Additionally, the increased difficulty of the prior shift test set exacerbates the loss gap between different experiments. Finally, the relation shift test set, characterized by its complex relationships and limited samples, results in slightly worse performance for the CAL model compared to the 1024-layer.

\section{Conclusion}
In materials property prediction, shifts in distribution are unavoidable and cause significant problems for standard machine learning models. Our investigation confirms that existing material property prediction models struggle with out-of-distribution (OOD) data. This limitation leads to unreliable predictions and hinders real-world applications in material science. To address this challenge, we explored the impact of various distributional shifts, including covariate shift (variations in input features), prior shift (variations in target labels), and relation shift (changes in the underlying relationship between inputs and targets). Our experiments showed that our CAL models generalize better overall, sometimes at the cost of performance on data similar to the training data. For small datasets such as the piezoelectric property prediction problem, CAL can adapt a base model more effectively than fine-tuning, the dominant paradigm in transfer learning.
We find that no single model outperforms the others on all datasets. As such, it is vital that materials property models are carefully evaluated on data that is as close as possible to the datasets to which they will be applied. 
Our work makes three key contributions. First, by constructing custom datasets exhibiting three different distribution shifts, we demonstrate the severity of OOD issues in current ML models for property prediction. This finding highlights a critical limitation that needs to be addressed for robust material property prediction. Second, we propose a novel approach for OOD material property prediction that integrates adversarial learning with training set ranking, which effectively enhances model robustness under diverse test scenarios marked by different input and target distributions. Finally, in our experiments simulating realistic prediction scenarios, we demonstrate the usefulness of our adversarial learning method over the baseline training and fine-tuning approach for OOD property prediction, especially when training with limited data. These results demonstrate the CAL model's capability to handle data scarcity, a common challenge encountered in material science research.
In conclusion, our research significantly advances the field of OOD methods for materials science. It offers immediate practical utility by providing a robust solution for overcoming OOD limitations typical in extrapolative material property prediction. Additionally, it opens doors for further exploration in this crucial area, paving the way for more reliable and generalizable material property prediction models.

\section{Data \& Code Availability}
The datasets generated in this work and code are available from the corresponding author upon request. The code will be made available in the future upon publication at \url{https://github.com/ramborz/CAL}.

\section{Contribution}
Conceptualization, J.H.; methodology, Q.L., J.H., N.M.; software, N.M., Q.L.; resources, J.H.; writing--original draft preparation, Q.L., J.H., N.M.; writing--review and editing, J.H., N.M., Q.L.; visualization, J.H., Q.L., N.M; supervision, J.H.; funding acquisition, J.H.

\section*{Acknowledgement}
The research reported in this work was supported in part by National Science Foundation under the grant 2110033, OAC-2311202, and 2320292. The views, perspectives, and content do not necessarily represent the official views of the NSF.

\bibliographystyle{unsrt}  
\bibliography{references}  

\begin{thebibliography}{10}

\bibitem{zagorac2019recent}
Dejan Zagorac, H~M{\"u}ller, S~Ruehl, J~Zagorac, and Silke Rehme.
\newblock Recent developments in the inorganic crystal structure database:
  theoretical crystal structure data and related features.
\newblock {\em Journal of applied crystallography}, 52(5):918--925, 2019.

\bibitem{allmann2007introduction}
Rudolf Allmann and Roland Hinek.
\newblock The introduction of structure types into the inorganic crystal
  structure database icsd.
\newblock {\em Acta Crystallographica Section A: Foundations of
  Crystallography}, 63(5):412--417, 2007.

\bibitem{belsky2002new}
Alec Belsky, Mariette Hellenbrandt, Vicky~Lynn Karen, and Peter Luksch.
\newblock New developments in the inorganic crystal structure database (icsd):
  accessibility in support of materials research and design.
\newblock {\em Acta Crystallographica Section B: Structural Science},
  58(3):364--369, 2002.

\bibitem{jain2013commentary}
Anubhav Jain, Shyue~Ping Ong, Geoffroy Hautier, Wei Chen, William~Davidson
  Richards, Stephen Dacek, Shreyas Cholia, Dan Gunter, David Skinner, Gerbrand
  Ceder, and A.~Persson Kristin.
\newblock Commentary: The materials project: A materials genome approach to
  accelerating materials innovation.
\newblock {\em APL materials}, 1(1):011002, 2013.

\bibitem{wang2021compositionally}
Anthony Yu-Tung Wang, Steven~K Kauwe, Ryan~J Murdock, and Taylor~D Sparks.
\newblock Compositionally restricted attention-based network for materials
  property predictions.
\newblock {\em Npj Computational Materials}, 7(1):77, 2021.

\bibitem{de2021materials}
Pierre-Paul De~Breuck, Geoffroy Hautier, and Gian-Marco Rignanese.
\newblock Materials property prediction for limited datasets enabled by feature
  selection and joint learning with modnet.
\newblock {\em npj Computational Materials}, 7(1):83, 2021.

\bibitem{chen2019graph}
Chi Chen, Weike Ye, Yunxing Zuo, Chen Zheng, and Shyue~Ping Ong.
\newblock Graph networks as a universal machine learning framework for
  molecules and crystals.
\newblock {\em Chemistry of Materials}, 31(9):3564--3572, 2019.

\bibitem{xie2018crystal}
Tian Xie and Jeffrey~C Grossman.
\newblock Crystal graph convolutional neural networks for an accurate and
  interpretable prediction of material properties.
\newblock {\em Physical review letters}, 120(14):145301, 2018.

\bibitem{jha2019irnet}
Dipendra Jha, Logan Ward, Zijiang Yang, Christopher Wolverton, Ian Foster,
  Wei-keng Liao, Alok Choudhary, and Ankit Agrawal.
\newblock Irnet: A general purpose deep residual regression framework for
  materials discovery.
\newblock In {\em Proceedings of the 25th ACM SIGKDD International Conference
  on Knowledge Discovery \& Data Mining}, pages 2385--2393, 2019.

\bibitem{hu2024realistic}
Jeffrey Hu, David Liu, Nihang Fu, and Rongzhi Dong.
\newblock Realistic material property prediction using domain adaptation based
  machine learning.
\newblock {\em Digital Discovery}, 3(2):300--312, 2024.

\bibitem{RealWorldMolecularOOD}
Prudencio Tossou, Cas Wognum, Michael Craig, Hadrien Mary, and Emmanuel
  Noutahi.
\newblock Real-world molecular out-of-distribution: Specification and
  investigation.
\newblock {\em Journal of Chemical Information and Modeling}, 64(3):697--711,
  2024.
\newblock PMID: 38300258.

\bibitem{lee2023MOOD}
Seul Lee, Jaehyeong Jo, and Sung~Ju Hwang.
\newblock Exploring chemical space with score-based out-of-distribution
  generation.
\newblock {\em Proceedings of the 40th International Conference on Machine
  Learning}, 2023.

\bibitem{yang2022learning}
Nianzu Yang, Kaipeng Zeng, Qitian Wu, Xiaosong Jia, and Junchi Yan.
\newblock Learning substructure invariance for out-of-distribution molecular
  representations.
\newblock In {\em Advances in Neural Information Processing Systems (NeurIPS)},
  2022.

\bibitem{omee2024structure}
Sadman~Sadeed Omee, Nihang Fu, Rongzhi Dong, Ming Hu, and Jianjun Hu.
\newblock Structure-based out-of-distribution (ood) materials property
  prediction: a benchmark study.
\newblock {\em npj Computational Materials}, 10(1):144, 2024.

\bibitem{xiong2020evaluating}
Zheng Xiong, Yuxin Cui, Zhonghao Liu, Yong Zhao, Ming Hu, and Jianjun Hu.
\newblock Evaluating explorative prediction power of machine learning
  algorithms for materials discovery using k-fold forward cross-validation.
\newblock {\em Computational Materials Science}, 171:109203, 2020.

\bibitem{loftis2020lattice}
Christian Loftis, Kunpeng Yuan, Yong Zhao, Ming Hu, and Jianjun Hu.
\newblock Lattice thermal conductivity prediction using symbolic regression and
  machine learning.
\newblock {\em The Journal of Physical Chemistry A}, 125(1):435--450, 2020.

\bibitem{back2024accelerated}
Seoin Back, Alan Aspuru-Guzik, Michele Ceriotti, Ganna Gryn'ova,
  Bartosz~Andrzej Grzybowski, Geun~Ho Gu, Jason~E Hein, Kedar Hippalgaonkar,
  Rodrigo Hormazabal, Yousung Jung, et~al.
\newblock Accelerated chemical science with ai.
\newblock {\em Digital Discovery}, 2024.

\bibitem{qi2023latent}
Han Qi, Stefano Rando, Xinyang Geng, Iku Ohama, Aviral Kumar, and Sergey
  Levine.
\newblock Latent conservative objective models for offline data-driven crystal
  structure prediction.
\newblock In {\em Workshop on ''Machine Learning for Materials'' ICLR 2023},
  2023.

\bibitem{Zou_2023_ICCV}
Yuli Zou, Weijian Deng, and Liang Zheng.
\newblock Adaptive calibrator ensemble: Navigating test set difficulty in
  out-of-distribution scenarios.
\newblock In {\em Proceedings of the IEEE/CVF International Conference on
  Computer Vision (ICCV)}, pages 19333--19342, October 2023.

\bibitem{magar2023learning}
Rishikesh Magar and Amir~Barati Farimani.
\newblock Learning from mistakes: Sampling strategies to efficiently train
  machine learning models for material property prediction.
\newblock {\em Computational Materials Science}, 224:112167, 2023.

\bibitem{moreno2012unifying}
Jose~G Moreno-Torres, Troy Raeder, Roc{\'\i}o Alaiz-Rodr{\'\i}guez, Nitesh~V
  Chawla, and Francisco Herrera.
\newblock A unifying view on dataset shift in classification.
\newblock {\em Pattern recognition}, 45(1):521--530, 2012.

\bibitem{liu2021stable}
Jiashuo Liu, Zheyan Shen, Peng Cui, Linjun Zhou, Kun Kuang, Bo~Li, and Yishi
  Lin.
\newblock Stable adversarial learning under distributional shifts.
\newblock In {\em Proceedings of the AAAI Conference on Artificial
  Intelligence}, volume~35, pages 8662--8670, 2021.

\bibitem{ward2018matminer}
Logan Ward, Alexander Dunn, Alireza Faghaninia, Nils~ER Zimmermann, Saurabh
  Bajaj, Qi~Wang, Joseph Montoya, Jiming Chen, Kyle Bystrom, Maxwell Dylla,
  et~al.
\newblock Matminer: An open source toolkit for materials data mining.
\newblock {\em Computational Materials Science}, 152:60--69, 2018.

\bibitem{ward2016general}
Logan Ward, Ankit Agrawal, Alok Choudhary, and Christopher Wolverton.
\newblock A general-purpose machine learning framework for predicting
  properties of inorganic materials.
\newblock {\em npj Computational Materials}, 2(1):1--7, 2016.

\bibitem{li2023global}
Qinyang Li, Rongzhi Dong, Nihang Fu, Sadman~Sadeed Omee, Lai Wei, and Jianjun
  Hu.
\newblock Global mapping of structures and properties of crystal materials.
\newblock {\em Journal of Chemical Information and Modeling}, 2023.

\bibitem{umap}
Leland McInnes, John Healy, Nathaniel Saul, and Lukas Großberger.
\newblock Umap: Uniform manifold approximation and projection.
\newblock {\em Journal of Open Source Software}, 3(29):861, 2018.

\bibitem{ema}
Boris Polyak and Anatoli~B. Juditsky.
\newblock Acceleration of stochastic approximation by averaging.
\newblock {\em Siam Journal on Control and Optimization}, 30:838--855, 1992.

\bibitem{layernorm}
Jimmy~Lei Ba, Jamie~Ryan Kiros, and Geoffrey~E. Hinton.
\newblock Layer normalization, 2016.

\bibitem{esfahani2015data}
Peyman~Mohajerin Esfahani and Daniel Kuhn.
\newblock Data-driven distributionally robust optimization using the
  wasserstein metric: Performance guarantees and tractable reformulations.
\newblock {\em arXiv preprint arXiv:1505.05116}, 2015.

\bibitem{arjovsky2019invariant}
Martin Arjovsky, L{\'e}on Bottou, Ishaan Gulrajani, and David Lopez-Paz.
\newblock Invariant risk minimization.
\newblock {\em arXiv preprint arXiv:1907.02893}, 2019.

\bibitem{dunn2020benchmarking}
Alexander Dunn, Qi~Wang, Alex Ganose, Daniel Dopp, and Anubhav Jain.
\newblock Benchmarking materials property prediction methods: the matbench test
  set and automatminer reference algorithm.
\newblock {\em npj Computational Materials}, 6(1):138, 2020.

\end{thebibliography}

\end{document}


\maketitle

\begin{table}[h!]
  \begin{center}
    \caption{OOD performance comparison: smoothed mean squared error for models predicting  predicting dielectirc refractive index (unit-less).
            This data set is challenging due the limited samples (4,764) and extreme target value distribution with 0.8085 median absolute deviation.
            Both our CAL baseline and CAL Finetune model achieves significantly better results compared to the IRnet models across all three OOD test sets.
    }
    
    \begin{threeparttable}[b]
    \begin{tabular}{r|r r r r r}
       \toprule %
                     &Train    & Test   & Covariate     & Maximum prior & Minimum prior \\
                     &  set    &  set   &   shift       &  shift        & shift \\
      \midrule %
        IR-net baseline       & 0.0458 & 0.0590 & 0.3030 & 2.4899 & 0.0823     \\
        Our CAL base          & \textbf{0.0122} & \textbf{0.0125} & 0.2451 & 2.3328 & 0.1194  \\
        \midrule %

        IR-net Fine-tune   &         &       & 0.1891 & 2.2243 & 0.0230 \\
        Our Targeted CAL   &          &      & \textbf{0.1823} & \textbf{1.7353} & 0.0284 \\
                \midrule
        Fine-tune \% improvement             &        &  & 37.58\% & 10.67\% & 71.99\% \\
        Targeted CAL \% improvement          &        &  &\textbf{ 25.62\%} & \textbf{25.61\%} & \textbf{76.21\%} \\
       \bottomrule %
    \end{tabular}
          \end{threeparttable}
  \label{tab:phonon}
  \end{center}
\end{table}

\begin{table}[h!]
  \begin{center}
    \caption{OOD performance comparison: smoothed mean squared error for models predicting  predicting base 10 logarithm of the DFT Voigt-Reuss-Hill average shear moduli in GPa.
            This data set had the 10987 samples and a 0.8085 median absolute deviation target value distribution.
    }
    
    \begin{threeparttable}[b]
    \begin{tabular}{r|r r r r r}
       \toprule %
                     &Train    & Test   & Covariate     & Maximum prior & Minimum prior \\
                     &  set    &  set   &   shift       &  shift        & shift \\
      \midrule %
        IR-net baseline  &    0.0049               & 0.0125               & 0.0339               & 0.0588               & 0.0837               \\
        Our CAL base      &    0.0082               & 0.0122               & 0.0402               & 0.0429               & 0.1156     \\
        \midrule %

        IR-net Fine-tune   &         &       & 0.0369 &0.0134 & 0.0575 \\
        Our Targeted CAL   &          &      & \textbf{0.0263} & \textbf{0.0122} & 0.0709 \\
                \midrule
        Fine-tune \% improvement             &        &  & -8.85\% & 77.15\% & 31.38\% \\
        Targeted CAL \% improvement          &        &  &\textbf{ 34.58\%} & 71.56\% & \textbf{38.67\%} \\
       \bottomrule %
    \end{tabular}
          \end{threeparttable}
  \label{tab:phonon}
  \end{center}
\end{table}